\documentstyle[preprint,aps]{revtex}
\begin{document}
\title{{\bf Exact Quantization Rule to the Kratzer-Type Potentials: An Application
to the Diatomic Molecules}}
\author{Sameer M. Ikhdair\thanks{%
sikhdair@neu.edu.tr} and \ Ramazan Sever\thanks{%
sever@metu.edu.tr}}
\address{$^{\ast }$Department of Physics, \ Near East University, Nicosia, Cyprus,
Mersin-10, Turkey\\
$^{\dagger }$Department of Physics, Middle East Technical \ University,
06531 Ankara, Turkey.}
\date{\today
}
\maketitle

\begin{abstract}
For any arbitrary values of $n$ and $l$ quantum numbers, we present a simple
exact analytical solution of the $D$-dimensional ($D\geq 2$) hyperradial Schr%
\"{o}dinger equation with the Kratzer and the modified Kratzer potentials
within the framework of the exact quantization rule (EQR) method. The exact
energy levels $(E_{nl})$ of all the bound-states are easily calculated from
this EQR method. The corresponding normalized hyperradial wave functions $%
(\psi _{nl}(r))$ are also calculated. The exact energy eigenvalues for these
Kratzer-type potentials are calculated numerically for the typical diatomic
molecules $LiH,$ $CH,$ $HCl,$ $CO,$ $NO,$ $O_{2},$ $N_{2}$ and $I_{2}$ for
various values of $n$ and $l$ quantum numbers. Numerical tests using the
energy calculations for the interdimensional degeneracy ($D=2-4$) for $I_{2},
$ $LiH,$ $HCl,$ $O_{2},$ $NO$ and $CO$ are also given. Our results obtained
by EQR are in exact agreement with those obtained by other methods.


PACS\ number: 03.65.-w; 03.65.Fd; 03.65.Ge.
\end{abstract}

\bigskip

% INITIALIZE - DONT CHANGE % %  %

\section{Introduction}

\noindent It is well-known that the exact analytical solution of the
hyperradial Schr\"{o}dinger equation, in any arbitrary spatial dimension $%
(D\geq 2),$ for its bound state energy levels, is fundamental to
understanding of molecular spectrum of a diatomic molecules in a
nonrelativistic quantum mechanics, since the resulting wave function
contains all the necessary information to describe a quantum system fully.
There are only a few potentials for which the radial Schr\"{o}dinger
equation can be solved explicitly for all $n$ and all $l$ quantum numbers$.$
So far, some useful methods have been developed, such as supersymmetry
(SUSY) [1], the Nikiforov-Uvarov method (NU) [2], the Pekris approximation
[3], asymptotic iteration method (AIM) [4], variational [5], the hypervirial
perturbation [6], the shifted $1/N$ expansion (SE) and the modified shifted $%
1/N$ expansion (MSE) [7], exact quantization rule (EQR) [8], perturbative
formalism [9], polynomial solution [10] and wave function ansatz method [11]
to solve the radial Schr\"{o}dinger equation exactly or quasi-exactly for $%
l\neq 0$ within a given potential.

The kratzer and modified Kratzer-type potentials [12] that we study in the
present work are important molecular potentials describe the interaction
between two atoms and have attracted a great of interest for some decades in
the history of quantum chemistry. Until now, these potentials have been used
extensively to describe the molecular structure and interactions [13]. In
this work, we attempt to apply the recently proposed EQR method [8], as an
alternative approach, to obtain analytically ($l\neq 0$) exact energy levels
and the corresponding wave functions of the $D$-dimensional $(D\geq 2)$
Schr\"{o}dinger equation with the Kratzer-type potentials. Further, we
calculate numerically these bound state energy levels for $LiH,$ $CH,$ $HCl,$
$CO,$ $NO,$ $O_{2},$ $N_{2}$ and $I_{2}$ diatomic molecules for different
values of $n$ and $l$ quantum numbers. Further, we give numerical tests
using energy calculations for interdimensional degeneracy, i.e., $%
(n,l,D)\rightarrow (n,l\pm 1,D\mp 2)$ corresponding to the confined $D=2-4$
diimensional Kratzer's potential.

This paper is organized as follows. In Section \ref{EQR} the EQR method is
reviewed. In Section \ref{SOTK}, the $D$-dimensional ($D\geq 2$)
Schr\"{o}dinger equation is solved by this method with $l\neq 0$ quantum
numbers to obtain the energy eigenvalues and the corresponding hyperradial
wave functions of the Kratzer-type potential. In Section \ref{ID} the
interdimensional degeneracy is introduced. Numerical calculations of the
energy levels and test of interdimensional degeneracy for many typical
diatomic molecules $CO,$ $NO,$ $CH,$ $HCl,$ $LiH,$ $N_{2},$ $O_{2},$ and $%
I_{2}$ are performed for various values of $n$ and $l$ quantum numbers to
compare with those obtained by other methods in Section \ref{NA}. Finally,
some concluding remarks are given in Section \ref{CR}.

\section{Exact Quantization Rule Method}

\label{EQR}A brief outline to the EQR method is presented with an extension
to the $D$-dimensional space ($D\geq 2$). The details can be found in Ref.
[8]. The EQR has recently been proposed to solve the one-dimensional ($1D$)
Schr\"{o}dinger equation:

\begin{equation}
\frac{d^{2}}{dx^{2}}\psi (x)+k(x)^{2}\psi (x)=0,\text{ \ }k(x)=\frac{1}{%
\hbar }\sqrt{2\mu \left[ E-V(x)\right] },
\end{equation}
where $\mu $ represents the reduced mass of the two interacting particles, $%
k(x)$ is the momentum and the potential $V(x)$ is a piecewise continuous
real function of $x.$ Equation (1) can be reduced into the Riccati equation:

\begin{equation}
\frac{d}{dx}\phi (x)+\phi (x)^{2}+k(x)^{2}=0,
\end{equation}
where $\phi (x)=\psi (x)^{-1}d\psi (x)/dx$ is the logarithmic derivative of
wave function $\psi (x).$ Due to the Sturm-Liouville theorem, the $\phi (x)$
decreases monotonically with respect to $x$ between two turning points,
where $E\geq V(x).$ Specifically, as $x$ increases across a node of the wave
function $\psi (x),$ where $E\geq V(x),$ $\phi (x)$ decreases to $-\infty ,$
jumps to $+\infty ,$ and then decreases again. By carefully studying the $1D$
Schr\"{o}dinger equation, Ma and Xu [8] obtained the EQR as

\begin{equation}
\int\limits_{x_{a}}^{x_{b}}k(x)dx=N\pi +\int\limits_{x_{a}}^{x_{b}}\phi (x)
\left[ \frac{dk(x)}{dx}\right] \left[ \frac{d\phi (x)}{dx}\right] ^{-1}dx,
\end{equation}
where $x_{a}$ and $x_{b}$ are two turning points determined by letting $%
E=V(x),$ $N$ is the number of nodes of \ $\phi (x)$ in the region $E\geq
V(x) $ and it is larger by one than the number of nodes of wave function $%
\psi (x),$ i.e. $N=n+1.$ Further, Ma and Xu [8] have generalized the EQR to $%
3D$ radial Schr\"{o}dinger equation with a spherically symmetric potential.
Here, we extend the $3D$-generalization of the EQR given by Ref. [8] into
the $D$-dimensional ($D\geq 2$) Schr\"{o}dinger equation. It is well-known
that in the $D$-dimensional hyperradial Schr\"{o}dinger equation, for any
arbitrary spherically symmetric potential, can be simply written as [14]
\[
\left[ \frac{d^{2}}{dr^{2}}+\frac{\left( D-1\right) }{r}\frac{d}{dr}-\frac{%
l(l+D-2)}{r^{2}}+\frac{2\mu }{\hbar ^{2}}\left( E-V(r)\right) \right] \psi
_{n,l,m}(r,\Omega _{D})=0,
\]
\begin{equation}
\psi _{n,l,m}(r,\Omega _{D})=\psi _{n,l}(r)Y_{l}^{m}(\Omega _{D}),
\end{equation}
where $l$ denotes the orbital angular momentum quantum number, $\mu $ is the
reduced mass, $\psi _{n,l}(r)$ is the hyperradial part of the wave function
and $Y_{l}^{m}(\Omega _{D})$ is the hyperspherical harmonics. Furthermore,
to remove the first derivative from the above equation, we define a new
radial wave function $R(r)$ by means of equation
\begin{equation}
\psi _{n,l}(r)=r^{-(D-1)/2}R(r),
\end{equation}
which will give the radial wave function $R(r)$ satisfying the wave equation
\begin{equation}
\left\{ \frac{d^{2}}{dr^{2}}-\frac{\left[ 4l\left( l+D-2\right) +\left(
D-1\right) \left( D-3\right) \right] }{4r^{2}}+\frac{2\mu }{\hbar ^{2}}%
\left( E-V(r)\right) \right\} R(r)=0.
\end{equation}
Further, the hyperradial equation (6) can be written into a simple analogy
of the $2D$ and $3D$ radial Schr\"{o}dinger equation given in Ref. [8] as
[15]
\begin{equation}
\frac{d^{2}}{dr^{2}}R(r)=-\frac{2\mu }{\hbar ^{2}}\left[ E-V_{\text{eff}}(r)%
\right] R(r),\text{ \ }V_{\text{eff}}(r)=\frac{\left( \eta ^{2}-1/4\right)
\hbar ^{2}}{2\mu r^{2}}+V(r),
\end{equation}
with the parameter
\begin{equation}
\eta =l+\frac{1}{2}\left( D-2\right) ,
\end{equation}
which is a linear combination of the spatial dimensions $D$ and the angular
momentum quantum number $l.$ Thus, the EQR can be generalized to the $D$%
-dimensional ($D\geq 2$) hyperradial Schr\"{o}dinger equation, with the new
effective potential form given by (7) as

\begin{equation}
\int\limits_{r_{a}}^{r_{b}}K(r)dr=N\pi +Q_{c},\text{ }K(r)=\frac{2\mu }{%
\hbar ^{2}}\left[ E-V_{\text{eff}}(r)\right] ,
\end{equation}
where
\begin{equation}
Q_{c}=\int\limits_{r_{a}}^{r_{b}}\phi (r)\left[ \frac{dk(r)}{dr}\right] %
\left[ \frac{d\phi (r)}{dr}\right] ^{-1}dr,
\end{equation}
is called the quantum correction, $N\pi $ is the contribution from the nodes
of wave function and $K(r)$ being the momentum between the two turning
points $r_{a}$ and $r_{b}.$ It has also been found that this quantum
correction is independent of the number of nodes of wave function for
solvable quantum system. Thus, the quantum correction in Eq. (10) can be
replaced by the ground-state calculations:
\begin{equation}
Q_{c}=\int\limits_{r_{a}}^{r_{b}}\phi _{0}(r)\left[ \frac{dk_{0}(r)}{dr}%
\right] \left[ \frac{d\phi _{0}(r)}{dr}\right] ^{-1}dr=\pi q.
\end{equation}

\section{Solutions of the Kratzer and Modified Kratzer-Type Potentials}

\label{SOTK}Considerable interest has recently been shown in the
Kratzer-Fues [12,16-19] as a model to describe internuclear vibration of a
diatomic molecule [20-23]. This potential can be expressed in the form
\begin{equation}
V(r)=-D_{e}\left( \frac{2r_{e}}{r}-\frac{r_{e}^{2}}{r^{2}}\right) ,
\end{equation}
where $D_{e}$ is the dissociation energy between two atoms in a solid and $%
r_{e}$ is the equilibrium internuclear length. This type of the potential in
(12) has been studied through the smooth transformation [16], the algebraic
approach [17], and the AIM [18]. The standard Kratzer's potential has been
recently modified by adding a $D_{e}$ term to the potential [24]. A new type
of this potential is called the modified Kratzer potential:

\begin{equation}
V(r)=D_{e}\left( \frac{r-r_{e}}{r}\right) ^{2},
\end{equation}
which is shifted in amount of $D_{e}.$ The above potential has been studied
in $D$-dimensions through the polynomial solutions [25] and also the
selection of a suitable wave function ansatz [15]. The modified Kratzer
potential has also been studied before in the $3D$ through the NU method
[24]. This potential can be simply taken as
\begin{equation}
V(r)=\frac{A}{r^{2}}-\frac{B}{r}+C,
\end{equation}
where $A=D_{e}r_{e}^{2},$ $B=2D_{e}r_{e}$ and $C=D_{e}$ [15,25,26]$.$ For
this spherically symmetric modified Kratzer potential, the effective
potential in Eq. (7) can be arranged to

\begin{equation}
V_{\text{eff}}(r)=\frac{2\mu A+\frac{1}{4}\left[ (M-2)^{2}-1\right] \hbar
^{2}}{2\mu r^{2}}-\frac{B}{r}+C,\text{ \ }M=D+2l.
\end{equation}
Further, by introducing the notation

\begin{equation}
\Lambda (\Lambda +1)\hbar ^{2}=2\mu A+\frac{1}{4}\left[ (M-2)^{2}-1\right]
\hbar ^{2},\text{ \ }
\end{equation}
from which we obtain a physically acceptable solution

\begin{equation}
\Lambda =\frac{-1+\sqrt{(M-2)^{2}+8\mu A/\hbar ^{2}}}{2},
\end{equation}
and finally the effective potential becomes

\begin{equation}
V_{\text{eff}}(r)=\frac{\Lambda (\Lambda +1)\hbar ^{2}}{2\mu r^{2}}-\frac{B}{%
r}+C.
\end{equation}
On the other hand, by taking $\phi (r)=R(r)^{-1}dR(r)/dr,$ the Riccati
equation reads

\begin{equation}
\frac{d}{dr}\phi (r)+\phi (r)^{2}=-\frac{2\mu }{\hbar ^{2}}\left[ E-C-\frac{%
\Lambda (\Lambda +1)\hbar ^{2}}{2\mu r^{2}}+\frac{B}{r}\right] .
\end{equation}
Now, we apply the EQR [8,26,27] given by (9) to solve the above Riccati
equation. At first, we must determine the turning points $r_{a}$ and $r_{b}$
from solving the relation $E=V_{\text{eff}}(r)$ with $r_{a}<$ $r_{b}.$ Thus,
we find

\begin{equation}
r_{a}=\frac{1}{-2(E-C)}\left[ B-\sqrt{B^{2}+\frac{2\Lambda (\Lambda
+1)(E-C)\hbar ^{2}}{\mu }}\right] ,
\end{equation}

\begin{equation}
r_{b}=\frac{1}{-2(E-C)}\left[ B+\sqrt{B^{2}+\frac{2\Lambda (\Lambda
+1)(E-C)\hbar ^{2}}{\mu }}\right] .
\end{equation}
The momentum between two turning points $r_{a}$ and $r_{b}$ can be
calculated by $k(r)=\sqrt{2\mu \left[ E_{0}-V_{\text{eff}}(r)\right] }/\hbar
$ as

\begin{equation}
k(r)=\frac{1}{\hbar r}\sqrt{-2\mu (E-C)(r-r_{a})(r_{b}-r)}.
\end{equation}
Since the phase angle $\phi (r)$ decreases as $r$ increases, then the
solution with one node and no pole only has a unique solution $\phi
_{0}(r)=\alpha _{1}r^{-1}+\alpha _{2}$ with $\alpha _{1}>0$ due to the
monotonic property$.$ Substituting this solution into the Riccati equation
(19), we find the ground-state solution as

\begin{equation}
\phi _{0}(r)=\frac{\Lambda +1}{r}-\frac{\mu B}{(\Lambda +1)\hbar ^{2}},\text{
with }E_{0}=C-\frac{\mu B^{2}}{2(\Lambda +1)^{2}\hbar ^{2}}.
\end{equation}
Evidently $\phi _{0}(r)$ is negative when $r\rightarrow \infty $, so that
the solution satisfies the physically admissible boundary condition. In
order to obtain the energy levels of all the bound-states, making use of the
integrals given in Appendix A, we are able to calculate the quantum
correction from Eq. (11) as

\[
Q_{c}=\frac{\sqrt{-2\mu (E_{0}-C)}}{2(\Lambda +1)^{2}\hbar ^{3}}%
\int\limits_{r_{a}}^{r_{b}}\frac{\left[ (\Lambda +1)^{2}\hbar ^{2}-\mu Br%
\right] }{r\sqrt{(r-r_{a})(r_{b}-r)}}\left[ (r_{a}+r_{b})r-2r_{a}r_{b}\right]
dr
\]

\begin{equation}
=\left[ \Lambda -\sqrt{\Lambda (\Lambda +1)}\right] \pi ,
\end{equation}
where the following formulas

\begin{equation}
r_{a}+r_{b}=-\frac{B}{(E_{0}-C)},\text{ \ }r_{a}r_{b}=-\frac{\Lambda
(\Lambda +1)\hbar ^{2}}{2\mu (E_{0}-C)},\text{ }\sqrt{-2\mu (E_{0}-C)}=\frac{%
\mu B}{(\Lambda +1)\hbar },
\end{equation}
have been used in finding Eq. (24). On the other hand, for any $n$ state, we
calculate the left hand side of Eq. (9) for $V_{\text{eff}}(r)$ given in Eq.
(18) using Eq. (22) as
\begin{equation}
\int\limits_{r_{a}}^{r_{b}}K(r)dr=\left[ \frac{\mu B}{\sqrt{-2\mu (E_{n}-C)}%
\hbar }-\sqrt{\Lambda (\Lambda +1)}\right] \pi .
\end{equation}
Further, the substitution of Eqs (24) and (26) into Eq. (9), enables us to
obtain the energy levels of all bound states in any arbitrary dimension $%
D\geq 2$ as

\[
E_{nl}^{(D)}=C-\left( \frac{2\mu }{\hbar ^{2}}B^{2}\right) \left[ 2n+1+\sqrt{%
(D+2l-2)^{2}+\frac{8\mu }{\hbar ^{2}}A}\right] ^{-2}
\]
\begin{equation}
=C-\frac{\mu }{2\hbar ^{2}}\left( \frac{B}{\widetilde{n}}\right) ^{2},\text{
}\widetilde{n}=N+\Lambda ,\text{ \ }n,l=0,1,2,\cdots
\end{equation}
where $N=n+1$ was used with $n$ is the number of nodes in the wave functions
$R(r).$

Therefore, for the $D$-dimensional Kratzer's potential ($C=0$), the energy
spectra become

\begin{equation}
E_{nl}^{(D)}=-\left( \frac{2\mu }{\hbar ^{2}}D_{e}^{2}r_{e}^{2}\right) \left[
n+\frac{1}{2}+\sqrt{\left( l+\frac{D}{2}-1\right) ^{2}+\frac{2\mu }{\hbar
^{2}}D_{e}r_{e}^{2}}\right] ^{-2},
\end{equation}
and also for the modified Kratzer-type ($C=D_{e}$) potential read

\begin{equation}
E_{nl}^{(D)}=D_{e}-\left( \frac{2\mu }{\hbar ^{2}}D_{e}^{2}r_{e}^{2}\right) %
\left[ n+\frac{1}{2}+\sqrt{\left( l+\frac{D}{2}-1\right) ^{2}+\frac{2\mu }{%
\hbar ^{2}}D_{e}r_{e}^{2}}\right] ^{-2},\text{ }E_{nl}^{(D)}<D_{e}.
\end{equation}
In what follows, we want to obtain the corresponding normalized hyperwave
function. At first, we define

\begin{equation}
\nu (\nu +D-2)=l(l+D-2)+\frac{2\mu A}{\hbar ^{2}},
\end{equation}

\begin{equation}
\rho =2\kappa r,\text{ }\tau =\frac{\mu B}{\hbar ^{2}\kappa },\text{ }\kappa
=\sqrt{-2\mu (E-C)/\hbar ^{2}},
\end{equation}
where for bound state case, states with $E<C$, we have real numbers for the
parameters $\kappa $ and $\tau .$ Thus, the hyperradial equation (4)
transforms to a standard form given by

\begin{equation}
\psi _{n,l}^{\prime \prime }(\rho )+\frac{\left( D-1\right) }{\rho }\psi
_{n,l}^{\prime }(\rho )+\left[ -\frac{1}{4}+\frac{\tau }{\rho }-\frac{\nu
(\nu +D-2)}{\rho ^{2}}\right] \psi _{n,l}(\rho )=0,
\end{equation}

\begin{equation}
\nu =-\frac{(D-2)}{2}+\frac{1}{2}\sqrt{\left( 2l+D-2\right) ^{2}+\frac{8\mu A%
}{\hbar ^{2}}},
\end{equation}
whose solutions are given as [19,25]

\begin{equation}
\psi _{n,l}(\rho )=N(n,l)\rho ^{\nu }e^{-\rho /2}{}_{1}F_{1}(-\tau +\nu +%
\frac{D-1}{2},2\nu +D-1;\rho )
\end{equation}
from which we can obtain the quantum condition as $-\tau +\nu +\frac{D-1}{2}%
=-n,$ $n=0,1,2,\cdots $ $[28,29].$ Furthermore, we can renormalize the bound
state solutions in (34) since these solutions are finite and convergent for
both $\rho =0$ $(r=0)$ and $\rho \rightarrow \infty .$

On the other hand, using the following relationship between the associated
Laguerre function and the hypergeometric function:

\begin{equation}
L_{n}^{\alpha }(z)=\frac{\Gamma (\alpha +n+1)}{n!\Gamma (a+1)}%
{}_{1}F_{1}(-n,\alpha +1;z),
\end{equation}
we may further rewrite the wave functions as

\begin{equation}
\psi _{n,l}(\rho )={\cal N}(n,l)\rho ^{\nu }e^{-\rho /2}{}L_{n}^{(2\nu
+D-2)}(\rho ),
\end{equation}
where ${\cal N}(n,l)$ is the normalizing factor to be determined by the
renormalization condition

\begin{equation}
\int\limits_{0}^{\infty }\left| \psi _{n,l}(\rho )\right| ^{2}r^{(D-1)}dr=1.
\end{equation}
To find the normalizing factor, let us review the generalized Coulomb-like
integral $J_{n,\alpha }^{(\beta )}(z)$ for noninteger $\alpha $ and $\beta $
derived in [30] as

\[
J_{n,\alpha }^{(\beta )}(z)=\int\limits_{0}^{\infty }e^{-z}z^{\alpha +\beta
} \left[ L_{n}^{(\alpha )}(z)\right] ^{2}dz=\frac{\Gamma (\alpha +n+1)}{%
\Gamma (n+1)}\sum\limits_{j=0}^{n}(-1)^{j}\frac{\Gamma (n-j-\beta )}{\Gamma
(-j-\beta )}
\]

\begin{equation}
\times \frac{\Gamma (\alpha +\beta +j+1)}{\Gamma (\alpha +j+1)}\frac{1}{%
\Gamma (j+1)\Gamma (n-j+1)},\text{ }{\cal R}\text{e}(\alpha +\beta +1)>0.
\end{equation}
In the present case, we have $\beta =1.$ This $J_{n,\alpha }^{(1)}$ has only
two nonzero contributions (for $j=n-1$ and $n$) in the sum (38) because of
the gamma functions of negative integers. The result $J_{n,\alpha }^{(1)}=%
\frac{(2n+\alpha +1)\Gamma (\alpha +n+1)}{\Gamma (n+1)}$ enables us to write
for our case

\begin{equation}
J_{n,2\nu +D-2}^{(1)}=\frac{(2\nu +2n+D-1)(2\nu +n+D-2)!}{n!},
\end{equation}
which gives the normalization factor

\[
{\cal N}(n,l)=\left[ \frac{n!(2\kappa )^{D}}{(2\nu +2n+D-1)(2\nu +n+D-2)!}%
\right] ^{1/2},
\]

\begin{equation}
\kappa =\frac{\mu B}{\hbar ^{2}(n+\Lambda +1)},\text{ }\Lambda =\nu +\frac{%
(D-3)}{2}.
\end{equation}
Therefore, we can finally obtain the re-normalized hyper wave function for
the modified Kratzer potential as

\begin{equation}
\psi _{n,l,m}(r,\Omega _{D})=\left[ \frac{n!}{(\nu _{1}+2n+1)(\nu _{1}+n)!}%
\right] ^{1/2}(2\kappa )^{\nu _{2}}r^{\nu }e^{-\kappa r}{}L_{n}^{(\nu
_{1})}(2\kappa r)Y_{l}^{m}(\Omega _{D}),
\end{equation}
with

\begin{equation}
2\kappa =\frac{8\mu D_{e}r_{e}}{\hbar ^{2}(2n+\nu _{1}+1)},\text{ }\nu _{1}=%
\sqrt{\left( 2l+D-2\right) ^{2}+\frac{8\mu }{\hbar ^{2}}D_{e}r_{e}}\text{ ,}
\end{equation}
\begin{equation}
\nu _{2}=\frac{\nu _{1}+2}{2},\text{ }\nu =\frac{\nu _{1}-(D-2)}{2}.
\end{equation}
which is found to be consistent with our previous findings in Ref. [25]
using the polynomial solution.

\section{Interdimensional Degeneracy}

\label{ID}From Eq. (27), it can be seen that two interdimensional states are
degenerate whenever [31]

\begin{equation}
(n,l,D)\rightarrow (n,l\pm 1,D\mp 2).
\end{equation}
Thus, a knowledge of $E_{nl}^{(D)}$ for $D=2$ and $D=3$ provides the
information necessary to find $E_{nl}^{(D)}$ for other higher dimensions.

For example, $E_{0,4}^{(2)}=E_{0,3}^{(4)}=E_{0,2}^{(6)}=E_{0,1}^{(8)}.$ This
is the same transformational invariance described for bound states of free
atoms and molecules [32,33] and demonstrates the existence of
interdimensional degeneracies among states of the confined Kratzer
potential. \

\section{Numerical Applications}

\label{NA}In this work, we have given numerical calculations for some
typical diatomic molecules to compare with those obtained by other numerical
exact methods given by the AIM [18] and the NU method [24]. In this regard,
we have calculated the bound state energy levels with Eq. (28) using the
Kratzer's potential for $CO,$ $NO,$ $HCl,$ $LiH,$ $O_{2}$ and $I_{2}$
diatomic molecules for selected $n,$ $l$ quantum numbers with parameter
values given in Tables 1. These parameters values are taken from Refs.
[34-37]. Therefore, we give the explicit values of Kratzer's energy levels
for different values of $n$ and $l$ in Tables 2-3. We further compare these
results with those obtained by the AIM [18]. In tables 4-5, we have also
tested the interdimensional degeneracy ($D=2-4$) using the energy
calculations for $CO,$ $NO,$ $HCl,$ $LiH,$ $O_{2},$ and $I_{2}$ diatomic
molecules. On the other hand, for the recently proposed modified Kratzer
potential, we have used the parameters for $CO,$ $NO,$ $CH$ and $N_{2}$
diatomic molecules given in Table 6 to compare with the NU method [24]. For
this case, in Tables 7-8, we give the energy levels for $CO,$ $NO,$ $CH$ and
$N_{2}$ diatomic molecules for various $n,$ $l$ quantum numbers by using Eq.
(29).

\section{Concluding Remarks}

\label{CR}In this study, we have used a different approach to find the $%
l\neq 0$ analytical and numerical solutions to the $D$-dimensional
hyperradial Schr\"{o}dinger equation with Kratzer-type potentials for
various diatomic molecules within the framework of the exact quantization
rule. For such potentials, the problem is simply reduced to a Coulomb
potential plus an inverse quadratic potential term. The exact eigensolutions
for this particular case have been obtained, in a similar way as the
Hydrogenic solutions [30]. Further, we have calculated the exact bound state
energy eigenvalues and the corresponding normalized hyper wave functions for
various diatomic molecules for any $l$ angular momentum quantum number bound
by an exactly solvable Kratzer-type potential. The advantage of the present
method is that it enables one to find the energy eigenvalues directly in a
simple way. The method presented in this work can be applied to find the
energy eigenvalues and the corresponding eigenfunctions of the
Schr\"{o}dinger equation within a given potential for various diatomic
molecules.

It is obvious that Eqs. (27) and (36) (with Eqs. (31) and (40)) reduce to
the well-known eigenvalues and eigenfunctions for the $3D$ Kratzer-Fues
potential when $D=3$ [38,39]. Further, they reduce to the Coulomb-like
solutions in $D$-dimensions if $A=C=0$ [40]. Similarly, for $B=2Z>0$ and $%
A=\lambda ,$ $Z$ and $\lambda $ are real constants [41].

\acknowledgments This research was partially supported by the Scientific and
Technical Research Council of Turkey. S. M. Ikhdair wishes to dedicate this
work to his family for their love and assistance.\newpage

\appendix

\section{Some helpful formulas}

Here we list some helpful integrals, which are not available in the integral
table, used in the present work to calculate the momentum integral and the
quantum correction terms [26,27]:\

\begin{equation}
\int\limits_{r_{a}}^{r_{b}}\frac{r}{\sqrt{(r-r_{a})(r_{b}-r)}}dr=\frac{\pi }{%
2}(r_{a}+r_{b}),
\end{equation}

\begin{equation}
\int\limits_{r_{a}}^{r_{b}}\frac{1}{r\sqrt{(r-r_{a})(r_{b}-r)}}dr=\frac{\pi
}{\sqrt{r_{a}r_{b}}},
\end{equation}
\begin{equation}
\int\limits_{r_{a}}^{r_{b}}\frac{1}{\sqrt{(r-r_{a})(r_{b}-r)}}dr=\pi ,
\end{equation}
\begin{equation}
\int\limits_{r_{a}}^{r_{b}}\frac{1}{r}\sqrt{(r-r_{a})(r_{b}-r)}dr=\left[
\frac{1}{2}(r_{a}+r_{b})-\sqrt{r_{a}r_{b}}\right] \pi .
\end{equation}

\baselineskip= 2\baselineskip% double space the text
%\end{document}
\bigskip

\begin{table}[tbp]
\caption{Reduced masses and spectroscopically determined properties of
various diatomic molecules in the ground electronic state.}
\begin{tabular}{lllllll}
Parameters\tablenotemark[1]\tablenotetext[1]{The data listed in this table
are taken from [34-37].} & $LiH$ & $I_{2}$ & O$_{2}$ & $HCl$ & $NO$ & $CO$
\\
\tableline$D_{0}$ $($in $eV)$ & $2.515283695$ & $1.581791863$ & $5.156658828$
& $4.619061175$ & $8.043782568$ & $10.84514471$ \\
$r_{0}$ $($in $A^{\circ })$ & $1.5956$ & $2.662$ & $1.208$ & $1.2746$ & $%
1.1508$ & $1.1282$ \\
$\mu $ (in amu) & $0.8801221$ & $63.45223502$ & $7.997457504$ & $0.9801045$
& $7.468441000$ & $6.860586000$%
\end{tabular}
\end{table}

\begin{table}[tbp]
\caption{Comparison of the energy levels, $E_{nl}$ (in $eV$)$,$
corresponding to the $3D$ Kratzer's potential for the various $n$ and $l$
quantum numbers for the diatomic molecules $LiH$, $I_{2},$ and $O_{2},$
where $\hbar c=1973.29$ $eV$ $A^{\circ }.$}
\begin{tabular}{lllllll}
$n$ & $l$ & $LiH$ [EQR] & $I_{2}$ [EQR] & $I_{2}$ [AIM] & $O_{2}$ [EQR] & $%
O_{2}$ [AIM] \\
\tableline$0$ & $0$ & $-2.467310304097$ & $-1.579082576525$ & $-1.579082577$
& $-5.126358620071$ & $-5.126358625$ \\
$1$ & $0$ & $-2.375819214406$ & $-1.573687150333$ & $-1.573687151$ & $%
-5.066641146718$ & $-5.066641151$ \\
& $1$ & $-2.374107972668$ & $-1.573677924919$ & $-1.573677925$ & $%
-5.066292321402$ & $-5.066292323$ \\
$2$ & $0$ & $-2.289324266253$ & $-1.568319329698$ & $-1.568319330$ & $%
-5.007961110233$ & $-5.007961116$ \\
& $1$ & $-2.287705602815$ & $-1.568310151445$ & $-1.568310152$ & $%
-5.007618327191$ & $-5.007618329$ \\
& $2$ & $-2.284475215373$ & $-1.568291795200$ & $-1.568291796$ & $%
-5.006932902380$ & $-5.006932904$ \\
$3$ & $0$ & $-2.207468200275$ & $-1.562978926616$ & $-1.562978927$ & $%
-4.950294618656$ & $-4.950294624$ \\
& $1$ & $-2.205935555783$ & $-1.562969795203$ & $-1.562969796$ & $%
-4.949957739138$ & $-4.949957740$ \\
& $2$ & $-2.202876749755$ & $-1.562951532697$ & $-1.562951533$ & $%
-4.949284118344$ & $-4.949284119$ \\
& $3$ & $-2.198304679122$ & $-1.562924139740$ & $-1.562924140$ & $%
-4.948274032620$ & $-4.948274034$ \\
$4$ & $0$ & $-2.129925128672$ & $-1.557665754681$ & $-1.557665755$ & $%
-4.893618463868$ & $-4.893618469$ \\
& $1$ & $-2.128472514560$ & $-1.557656669790$ & $-1.557656670$ & $%
-4.893287353086$ & $-4.893287355$ \\
& $2$ & $-2.125573350591$ & $-1.557638500327$ & $-1.557638501$ & $%
-4.892625266816$ & $-4.892625268$ \\
& $3$ & $-2.121239701434$ & $-1.557611246928$ & $-1.557611247$ & $%
-4.891632475505$ & $-4.891632476$ \\
& $4$ & $-2.115489505754$ & $-1.557574910549$ & $-1.557574911$ & $%
-4.890309384483$ & $-4.890309388$ \\
$5$ & $0$ & $-2.056397286593$ & $-1.552379629069$ & $-1.552379630$ & $%
-4.837910098245$ & $-4.837910103$ \\
& $1$ & $-2.055019226505$ & $-1.552370590385$ & $-1.552370591$ & $%
-4.837584625235$ & $-4.837584627$ \\
& $2$ & $-2.052268785922$ & $-1.552352513333$ & $-1.552352514$ & $%
-4.836933811639$ & $-4.836933812$ \\
& $3$ & $-2.048157264859$ & $-1.552325398546$ & $-1.552325399$ & $%
-4.835957922172$ & $-4.835957923$ \\
& $4$ & $-2.042701466576$ & $-1.552289246974$ & $-1.552289248$ & $%
-4.834657353568$ & $-4.834657357$ \\
& $5$ & $-2.035923524667$ & $-1.552244059882$ & $-1.552244060$ & $%
-4.833032634174$ & $-4.833032637$%
\end{tabular}
\end{table}
\begin{table}[tbp]
\caption{Comparison of the energy levels, $E_{nl}$ (in $eV$)$,$
corresponding to the $3D$ Kratzer's potential for the various $n$ and $l$
quantum numbers for the diatomic molecules $HCl$, $NO,$ and $CO,$ where $%
\hbar c=1973.29$ $eV$ $A^{\circ }.$}
\begin{tabular}{lllllll}
$n$ & $l$ & $HCl$ [EQR] & $NO$ [EQR] & $NO[AIM]$ & $CO$ [EQR] & $CO[AIM]$ \\
\tableline$0$ & $0$ & $-4.541848211101$ & $-8.002659419493$ & $-8.002659417$
& $-10.794315323$ & $-10.79431532$ \\
$1$ & $0$ & $-4.393727956046$ & $-7.921456840689$ & $-7.921456839$ & $%
-10.693839925$ & $-10.69383992$ \\
& $1$ & $-4.391293850595$ & $-7.921043829925$ & $-7.921043834$ & $%
-10.693371229$ & $-10.69337123$ \\
$2$ & $0$ & $-4.252737112329$ & $-7.841483958093$ & $-7.841483956$ & $%
-10.594760890$ & $-10.59476089$ \\
& $1$ & $-4.250419208735$ & $-7.841077185904$ & $-7.841077188$ & $%
-10.594298692$ & $-10.59429869$ \\
& $2$ & $-4.245791052967$ & $-7.840263768523$ & $-7.840263771$ & $%
-10.593374417$ & $-10.59337441$ \\
$3$ & $0$ & $-4.118425371585$ & $-7.762716067159$ & $-7.762716066$ & $%
-10.497052462$ & $-10.49705246$ \\
& $1$ & $-4.116216389518$ & $-7.762315408528$ & $-7.762315413$ & $%
-10.496596643$ & $-10.49659664$ \\
& $2$ & $-4.111805631214$ & $-7.761514215884$ & $-7.761514218$ & $%
-10.495685124$ & $-10.49568512$ \\
& $3$ & $-4.105207449232$ & $-7.760312738370$ & $-7.760312744$ & $%
-10.494318144$ & $-10.49431814$ \\
$4$ & $0$ & $-3.990377425087$ & $-7.685129080626$ & $-7.685129079$ & $%
-10.400689478$ & $-10.40068947$ \\
& $1$ & $-3.988270645562$ & $-7.684734413653$ & $-7.684734417$ & $%
-10.400239921$ & $-10.40023992$ \\
& $2$ & $-3.984063879462$ & $-7.683945202003$ & $-7.683945203$ & $%
-10.399340924$ & $-10.39934092$ \\
& $3$ & $-3.977770657506$ & $-7.682761690175$ & $-7.682761696$ & $%
-10.397992722$ & $-10.39799272$ \\
& $4$ & $-3.969411138650$ & $-7.681184244677$ & $-7.681184246$ & $%
-10.396195666$ & $-10.39619567$ \\
$5$ & $0$ & $-3.868209749404$ & $-7.608699510108$ & $-7.608699509$ & $%
-10.305647347$ & $-10.30564735$ \\
& $1$ & $-3.866198963636$ & $-7.608310715917$ & $-7.608310719$ & $%
-10.305203938$ & $-10.30520394$ \\
& $2$ & $-3.862183802008$ & $-7.607533247563$ & $-7.607533248$ & $%
-10.304317236$ & $-10.30431723$ \\
& $3$ & $-3.856177032746$ & $-7.606367345012$ & $-7.606367349$ & $%
-10.302987469$ & $-10.30298747$ \\
& $4$ & $-3.848197679994$ & $-7.604813367976$ & $-7.604813368$ & $%
-10.301214985$ & $-10.30121499$ \\
& $5$ & $-3.838270872139$ & $-7.602871795644$ & $-7.602871795$ & $%
-10.299000242$ & $-10.29900024$%
\end{tabular}
\end{table}
\begin{table}[tbp]
\caption{Numerical tests using the energy calculations (in $eV$) for the
interdimensional degeneracy corresponding to the confined $D=2-4$
dimensional Kratzer potential for various $n,l$ quantum numbers for the
diatomic molecules $I_{2}$, $LiH,$ and $HCl,$ where $\hbar c=1973.29$ $eV$ $%
A^{\circ }.$}
\begin{tabular}{llllll}
$D$ & $l$ & $n$ & $I_{2}$ [EQR] & $LiH$ [EQR] & $HCl$ [EQR] \\
\tableline$2$ & $0$ & $0$ & $-1.579083735645$ & $-2.467536868725$ & $%
-4.542168190064$ \\
&  & $1$ & $-1.573688303518$ & $-2.376033294347$ & $-4.394032410177$ \\
&  & $2$ & $-1.568320476987$ & $-2.289526762193$ & $-4.253027029918$ \\
& $1$ & $0$ & $-1.579079099175$ & $-2.466630860376$ & $-4.540888545222$ \\
&  & $1$ & $-1.573683690790$ & $-2.375177207655$ & $-4.392814848343$ \\
&  & $2$ & $-1.568315887840$ & $-2.288716995880$ & $-4.251867599163$ \\
& $2$ & $0$ & $-1.579065189928$ & $-2.463916832438$ & $-4.537053942242$ \\
&  & $1$ & $-1.573669852770$ & $-2.372612671616$ & $-4.389166233568$ \\
&  & $2$ & $-1.568302120561$ & $-2.286291171344$ & $-4.248393136578$ \\
& 3 & $0$ & $-1.579042008394$ & $-2.459406732177$ & $-4.530677339114$ \\
&  & $1$ & $-1.573646789943$ & $-2.368350817764$ & $-4.383098743943$ \\
&  & $2$ & $-1.568279175633$ & $-2.282259674365$ & $-4.242615099462$ \\
$3$ & 0 & $0$ & $-1.579082576525$ & $-2.467310304097$ & $-4.541848211101$ \\
&  & $1$ & $-1.573687150333$ & $-2.375819214406$ & $-4.393727956046$ \\
&  & $2$ & $-1.568319329698$ & $-2.289324266254$ & $-4.252737112328$ \\
& 1 & $0$ & $-1.579073303626$ & $-2.465499287366$ & $-4.539290004878$ \\
&  & $1$ & $-1.573677924919$ & $-2.374107972668$ & $-4.391293850595$ \\
&  & 2 & $-1.568310151445$ & $-2.287705602815$ & $-4.250419208735$ \\
& 2 & 0 & $-1.579054758153$ & $-2.461885237093$ & $-4.534182246350$ \\
&  & 1 & $-1.573659474414$ & $-2.370692927106$ & $-4.386433772612$ \\
$4$ & 0 & $0$ & $-1.579079099175$ & $-2.466630860376$ & $-4.540888545222$ \\
&  & $1$ & $-1.573683690790$ & $-2.375177207655$ & $-4.392814848343$ \\
&  & 2 & $-1.568315887840$ & $-2.288716995880$ & $-4.251867599163$ \\
& 1 & 0 & $-1.579065189928$ & $-2.463916832438$ & $-4.537053942242$ \\
&  & 1 & $-1.573669852770$ & $-2.372612671616$ & $-4.389166233568$ \\
&  & 2 & $-1.568302120561$ & $-2.286291171344$ & $-4.248393136578$ \\
& 2 & 0 & $-1.579042008394$ & $-2.459406732177$ & $-4.530677339114$ \\
&  & 1 & $-1.573646789943$ & $-2.368350817764$ & $-4.383098743943$ \\
&  & 2 & $-1.568279175633$ & $-2.282259674365$ & $-4.242615099462$%
\end{tabular}
\end{table}
\begin{table}[tbp]
\caption{Numerical tests using the energy calculations (in $eV$) for the
interdimensional degeneracy corresponding to the confined $D=2-4$
dimensional Kratzer potential for various $n,l$ quantum numbers for the
diatomic molecules $O_{2}$, $NO,$ and $CO,$ where $\hbar c=1973.29$ $eV$ $%
A^{\circ }.$}
\begin{tabular}{llllll}
$D$ & $l$ & $n$ & $O_{2}$ [EQR] & $NO$ [EQR] & $CO$ [EQR] \\
\tableline$2$ & $0$ & $0$ & $-5.126402999836$ & $-8.002711844782$ & $%
-10.794374740920$ \\
&  & $1$ & $-5.066684753267$ & $-7.921508470069$ & $-10.693898515189$ \\
&  & $2$ & $-5.008003961426$ & $-7.841534807595$ & $-10.594818667644$ \\
& $1$ & $0$ & $-5.126225485390$ & $-8.002502147751$ & $-10.794137074251$ \\
&  & $1$ & $-5.066510331583$ & $-7.921301956595$ & $-10.693664159142$ \\
&  & $2$ & $-5.007832561073$ & $-7.841331413560$ & $-10.594587561016$ \\
& $2$ & $0$ & $-5.125693015861$ & $-8.001873122635$ & $-10.793424137068$ \\
&  & $1$ & $-5.065987138736$ & $-7.920682480901$ & $-10.692961152734$ \\
&  & $2$ & $-5.007318430661$ & $-7.840721294966$ & $-10.593894301796$ \\
& 3 & $0$ & $-5.124805812601$ & $-8.000824967303$ & $-10.792236117812$ \\
&  & $1$ & $-5.065115391268$ & $-7.919650237110$ & $-10.691789681130$ \\
&  & $2$ & $-5.006461782057$ & $-7.839704642277$ & $-10.592739071948$ \\
$3$ & 0 & $0$ & $-5.126358620071$ & $-8.002659419493$ & $-10.794315323271$
\\
&  & $1$ & $-5.066641146718$ & $-7.921456840689$ & $-10.693839925213$ \\
&  & $2$ & $-5.007961110233$ & $-7.841483958093$ & $-10.594760890040$ \\
& 1 & $0$ & $-5.126003609633$ & $-8.002240041928$ & $-10.793840005639$ \\
&  & $1$ & $-5.066292321402$ & $-7.921043829925$ & $-10.693371228552$ \\
&  & 2 & $-5.007618327191$ & $-7.841077185904$ & $-10.594298691949$ \\
& 2 & 0 & $-5.125293736355$ & $-8.001401418731$ & $-10.792889496018$ \\
&  & 1 & $-5.065594815162$ & $-7.920217937834$ & $-10.692433958691$ \\
$4$ & 0 & $0$ & $-5.126225485390$ & $-8.002502147751$ & $-10.794137074251$
\\
&  & $1$ & $-5.066510331583$ & $-7.921301956595$ & $-10.693664159142$ \\
&  & 2 & $-5.007832561073$ & $-7.841331413560$ & $-10.594587561016$ \\
& 1 & 0 & $-5.125693015861$ & $-8.001873122635$ & $-10.793424137068$ \\
&  & 1 & $-5.065987138736$ & $-7.920682480901$ & $-10.692961152734$ \\
&  & 2 & $-5.007318430661$ & $-7.840721294966$ & $-10.593894301796$ \\
& 2 & 0 & $-5.124805812601$ & $-8.000824967303$ & $-10.792236117812$ \\
&  & 1 & $-5.065115391268$ & $-7.919650237110$ & $-10.691789681130$ \\
&  & 2 & $-5.006461782057$ & $-7.839704642277$ & $-10.592739071948$%
\end{tabular}
\end{table}

\begin{table}[tbp]
\caption{Reduced masses and spectroscopically determined properties of $%
N_{2},$ $CO,$ $NO$ and $CH$ diatomic molecules in the ground electronic
state.}
\begin{tabular}{lllll}
Parameters\tablenotemark[1]\tablenotetext[1]{The data listed in this table
are taken from [24].} & $N_{2}$ & $CO$ & $NO$ & $CH$ \\
\tableline$D_{0}$ $(cm^{-1})$ & $96288.03528$ & $87471.42567$ & $64877.06229$
& $31838.08149$ \\
$r_{0}$ $(A^{\circ })$ & $1.0940$ & $1.1282$ & $1.1508$ & $1.1198$ \\
$\mu $ (amu) & $7.00335$ & $6.860586$ & $7.468441$ & $0.929931$%
\end{tabular}
\end{table}

\begin{table}[tbp]
\caption{Comparison of the energy eigenvalues, $E_{nl}$ (in $eV$)$,$
corresponding to the $3D$ modified Kratzer potential for the various $n$ and
$l$ quantum numbers for the diatomic molecules $N_{2}$ and $CO,$ where $%
\hbar c=1973.29$ $eV$ $A^{\circ }.$}
\begin{tabular}{llllll}
$n$ & $l$ & $N_{2}$ [EQR] & $N_{2}$ [NU] & $CO$ [EQR] & $CO$ [NU] \\
\tableline$0$ & $0$ & $0.054436738370$ & $0.054430$ & $0.050829386733$ & $%
0.050823$ \\
$1$ & $0$ & $0.162076974947$ & $0.162057$ & $0.151304784801$ & $0.151287$ \\
& $1$ & $0.162565552668$ & $0.162546$ & $0.151773481462$ & $0.151755$ \\
$2$ & $0$ & $0.268261347644$ & $0.268229$ & $0.250383819984$ & $0.250354$ \\
& $1$ & $0.268743332192$ & $0.268711$ & $0.250846018075$ & $0.250816$ \\
& $2$ & $0.269707181517$ & $0.269675$ & $0.251770292931$ & $0.251744$ \\
$3$ & $0$ & $0.373015993195$ & $0.372972$ & $0.348092247640$ & $0.348051$ \\
& $1$ & $0.373491502668$ & $0.373447$ & $0.348548066746$ & $0.348507$ \\
& $2$ & $0.374442403850$ & $0.374398$ & 0.349459585720 & $0.349418$ \\
& $3$ & $0.375868461286$ & $0.375823$ & 0.350826566166 & $0.350785$ \\
$4$ & $0$ & $0.476366464424$ & $0.476313$ & 0.444455232045 & $0.444403$ \\
& $1$ & $0.476835614288$ & $0.476779$ & 0.444904789014 & $0.444852$ \\
& $2$ & $0.477773798215$ & $0.477717$ & 0.445803785755 & $0.445751$ \\
& $3$ & $0.479180784677$ & $0.479124$ & 0.447151987956 & $0.447099$ \\
& $4$ & $0.481056226559$ & $0.480999$ & 0.448949044337 & $0.448895$ \\
$5$ & $0$ & $0.578337745829$ & $0.578269$ & 0.539497362596 & $0.539434$ \\
& $1$ & $0.578800648986$ & $0.578732$ & 0.539940771611 & $0.539877$ \\
& $2$ & $0.579726341424$ & $0.579658$ & 0.540827474443 & $0.540764$ \\
& $3$ & $0.581114595455$ & $0.581046$ & 0.542157240772 & $0.542093$ \\
& $4$ & $0.582965069724$ & $0.582896$ & 0543929725307 & $0.543865$ \\
& $5$ & $0.585277309425$ & $0.585208$ & 0.546144468004 & $0.546082$%
\end{tabular}
\end{table}

\begin{table}[tbp]
\caption{Comparison of the energy eigenvalues, $E_{nl}$ (in $eV$)$,$
corresponding to the $3D$ modified Kratzer potential for the various $n$ and
$l$ quantum numbers for the diatomic molecules $NO$ and $CH,$ where $\hbar
c=1973.29$ $eV$ $A^{\circ }.$}
\begin{tabular}{llllll}
$n$ & $l$ & $NO$ [EQR] & $NO$ [NU] & $CH$ [EQR] & $CH$ [NU] \\
\tableline$0$ & $0$ & $0.041123148507$ & $0.041118$ & $0.083224106534$ & $%
0.083214$ \\
$1$ & $0$ & $0.122325727312$ & $0.122311$ & $0.241151342060$ & $0.241123$ \\
& $1$ & $0.122738738076$ & $0.122724$ & $0.244409651454$ & $0.244381$ \\
$2$ & $0$ & $0.202298609907$ & $0.202274$ & $0.389591252689$ & $0.389547$ \\
& $1$ & $0.202705382097$ & $0.202681$ & $0.392655829482$ & $0.392611$ \\
& $2$ & $0.203518799478$ & $0.203494$ & $0.398768962343$ & $0.398722$ \\
$3$ & $0$ & $0.281066500842$ & $0.281033$ & $0.529288818460$ & $0.529229$ \\
& $1$ & $0.281467159473$ & $0.281434$ & $0.532174721830$ & $0.532115$ \\
& $2$ & $0.282268352117$ & $0.282235$ & $0.537931662876$ & $0.537870$ \\
& $3$ & $0.283469829631$ & $0.283436$ & $0.546530102918$ & $0.546467$ \\
$4$ & $0$ & $0.358653487375$ & $0.358611$ & $0.660917306208$ & $0.660844$ \\
& $1$ & $0.359048154348$ & $0.359006$ & $0.663638151412$ & $0.663565$ \\
& $2$ & $0.359837365998$ & $0.359795$ & $0.669066052612$ & $0.668992$ \\
& $3$ & $0.361020877826$ & $0.360978$ & $0.677173530261$ & $0.677098$ \\
& $4$ & $0.362598323324$ & $0.362555$ & $0.687919851051$ & $0.687842$ \\
$5$ & $0$ & $0.435083057893$ & $0.435032$ & $0.785086396757$ & $0.785001$ \\
& $1$ & $0.435471852084$ & $0.435421$ & $0.787654550920$ & $0.787569$ \\
& $2$ & $0.436249320438$ & $0.436198$ & $0.792778001145$ & $0.792692$ \\
& $3$ & $0.437415222989$ & $0.437364$ & $0.800431194100$ & $0.800343$ \\
& $4$ & $0.438969200025$ & $0.438917$ & $0.810576203106$ & $0.810487$ \\
& $5$ & $0.440910772357$ & $0.440858$ & $0.823163201411$ & $0.823071$%
\end{tabular}
\end{table}

\end{document}